\documentclass[final,3p,times,twocolumn]{elsarticle}
\usepackage{hyperref}
\hypersetup{  
	colorlinks=true, 
	linkcolor=blue, 
	anchorcolor=red, 
	urlcolor=cyan,   
	citecolor=green  
}
\usepackage{nameref} 
\usepackage{amssymb} 
\usepackage{amsmath} 
\usepackage{subcaption}
\usepackage{booktabs} 
\usepackage{multirow}
\usepackage{graphicx}
\usepackage{array}
\usepackage{xcolor}

\begin{document}
\captionsetup[figure]{labelformat=empty,textformat=period}
\begin{frontmatter}  
	\title{Brain–Muscle Atlas: A novel framework for Motor Brain–Computer Interfaces\tnoteref{label1}} 
	
	\tnotetext[label1]{This work is supported by the National Natural Science Foundation of China (62303423, 62373295), the STI 2030-Major Project (2022ZD0208500), Postdoctoral Science Foundation of China (2024T170844, 2023M733245), the Henan Province key research and development and promotion of special projects (242102311239), Shaanxi Provincial Key Research and Development Program (2023GXLH-012).} 
	
\author[1,2]{Ye Sun\textsuperscript{\dag}}
\ead{ye@gs.zzu.edu.cn}

\author[1]{Bowei Zhao\textsuperscript{\dag}}
\ead{15225355605@163.com}

\author[3,4,1,2]{Dezhong Yao\textsuperscript{\dag}}
\ead{dyao@uestc.edu.cn}

\author[1,2]{Rui Zhang\textsuperscript{\dag}}
\ead{ruizhang@zzu.edu.cn}

\author[5]{Bohan Zhang}
\ead{z5465167@ad.unsw.edu.au}

\author[1,2]{Xiaoyuan Li}
\ead{lixiaoyuan@zzu.edu.cn}

\author[6]{Jing Wang}
\ead{pele.wang@xjtu.edu.cn}
\author[1]{Mingxuan Qu}
\ead{13940369751@163.com}

\author[1,2]{Gang Liu\corref{cor1}}
\ead{gangliu_@zzu.edu.cn}

\address[1]{School of Electrical and Information Engineering, Zhengzhou University, Zhengzhou 450001, China}
\address[2]{Henan Provincial Key Laboratory of Brain Science and Brain-Computer Interface Technology, Henan, China}
\address[3]{Clinical Hospital of Chengdu Brain Science Institute, MOE-Key Lab for NeuroInformation, Brain-Apparatus Communication Institute, University of Electronic Science and Technology of China, Chengdu, China}
\address[4]{Research Unit of NeuroInformation 2019RU035, Chinese Academy of Medical Sciences, Chengdu 611731, China}
\address[5]{School of Computer Science and Engineering, University of New South Wales, Sydney NSW 2052, Australia}
\address[6]{School of Mechanical Engineering, Xi’an Jiaotong University, Xi’an 710049, China}

\cortext[cor1]{Corresponding author:Gang Liu}

	\journal{EXPERT SYSTEMS WITH APPLICATIONS}		
		\begin{abstract}
		Motor brain–computer interfaces (BCIs) enable the control of external devices by decoding neural signals. However, most existing systems rely on a direct “brain-machine” mapping, overlooking the hierarchical physiological pathway of natural movement, namely the “brain–muscle–joint” cascade. Due to the lack of explicit modeling and enhancement of this pathway, current systems are often constrained by the low amplitude and high noise of EEG signals, resulting in motor outputs that are unstable, discontinuous, and insufficiently natural. To address these limitations, this study introduces the concept of a brain–muscle atlas, designed to systematically characterize the mapping between motor cortical activity and corresponding muscle activation, thereby establishing a movement decoding framework that better aligns with neuromuscular physiology. Using synchronously recorded EEG–EMG data, we constructed the first brain–muscle atlas for elbow flexion–extension, achieving a structured mapping from cortical activity to muscle activation. Offline experiments demonstrate that the proposed atlas accurately reconstructs the temporal activation patterns of primary elbow agonists, achieving a maximum correlation coefficient of 0.8314, thereby validating its ability to capture cortical–muscular mapping. Furthermore, by leveraging atlas-derived muscle activation representations, we enabled continuous real-time control of a virtual elbow joint. All ten participants successfully completed the online flexion–extension task, indicating that the system robustly extracts motor intent even under low-SNR EEG conditions.
			
		\end{abstract}
	\begin{highlights}
		\item This paper proposes a brain–muscle atlas that reconstructs physiologically plausible cortico–myographic mappings grounded in the hierarchical “brain–muscle–joint’’ motor pathway.                         
		\item The resulting brain–muscle–elbow interface demonstrates a systems-level shift from direct “brain-to-machine’’ decoding toward biologically structured motor computation.          
           
	\end{highlights}
		
		\begin{keyword}
			 Brain–Muscle Atlas \sep Brain-muscle-elbow interface \sep Muscle function replacement
			
		\end{keyword}	
 	\end{frontmatter}
 
	\section{Introduction}
	\label{Introduction}

	Motor brain-computer interfaces (BCIs) aim to restore control capabilities for patients with motor dysfunction by interpreting neural activity~\cite{0,00}. However, mainstream non-invasive BCIs possess a fundamental flaw in their system architecture: they bypass the crucial "brain-muscle-joint" pathway intrinsic to natural motor control, instead establishing a direct mapping from the cortex to external devices~\cite{1,2}. This "physiology-bypassing" design, while simplifying the control loop, discards the optimized signal transmission and transformation mechanisms within the motor system. This results in crude, discontinuous decoded commands and severe instability in complex environments~\cite{3}.
	
	In natural movement, motor control relies on efficient hierarchical transmission between the cortex and muscles~\cite{4,5,5.1,5.2}. The synergistic activation of muscles serves as a critical "physiological relay," converting abstract cortical commands into execution-level directives with high signal-to-noise ratios, thereby enabling smooth and precise joint motion~\cite{6}. Traditional BCIs discard this relay, effectively bypassing several key functions implemented by the spinal-muscular system, including sensorimotor integration, signal gain modulation, and the spatiotemporal muscle activation patterns required for coordinated and efficient movement~\cite{5.2,7}.
	
	\begin{figure}[htpb]  
		\centering  
		\includegraphics[width=0.49\textwidth]{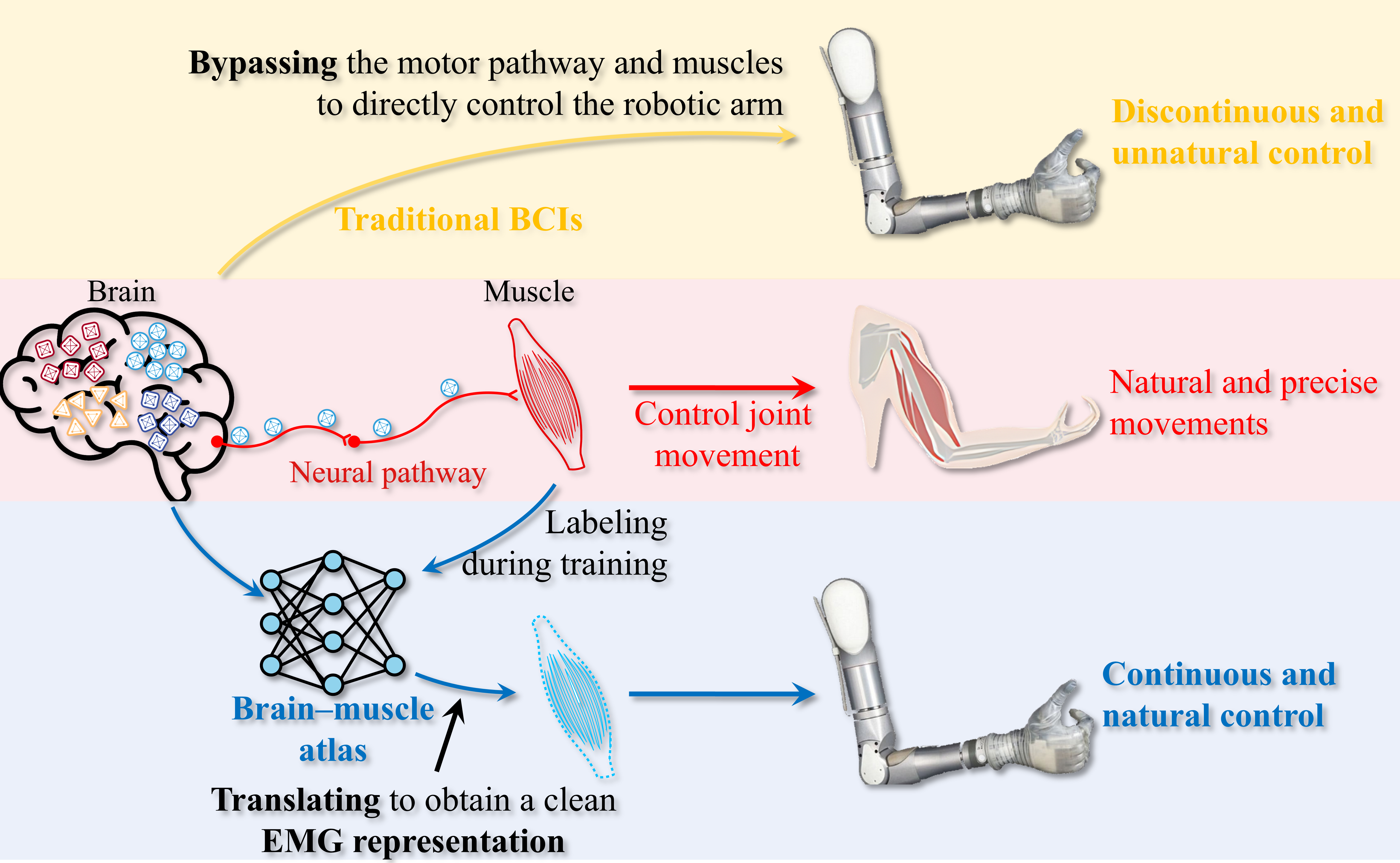} 
		\caption{\textbf{Figure 1.} The concept of the Brain-Muscle Atlas and the corresponding architecture of the Brain-Muscle-Elbow Interface} 
		\label{fig:fig1}
	\end{figure} 
	
	To address this bottleneck, we introduce a computable neuromuscular relay model as a novel computational paradigm for BCIs, designed to emulate the signal-transformation role of muscles within the natural motor pathway~\cite{8,9}. This paradigm is instantiated as a brain–muscle atlas, a dynamic mapping model that quantitatively characterises the full pathway from cortical activity to muscle activation.
	
	\autoref{fig:fig1} illustrates the concept of the proposed Brain-Muscle Atlas (BMA) and the corresponding architecture of the Brain-Muscle-Elbow Interface (BMEI). Traditional BCIs attempt to decode motor commands directly from electroencephalography (EEG) signals. However, they face fundamental challenges: due to skull-induced attenuation and aliasing of electrical signals, EEG is characterized by low amplitude, poor signal-to-noise ratio (SNR), and contains substantial neural activity unrelated to the target movement, resulting in severely insufficient movement specificity. In contrast, EMG signals, generated by muscle fiber discharges, exhibit millivolt-level potential changes measurable on the body surface. They offer significantly higher SNR and have activation patterns that correspond unambiguously to movement intention.
	
	Leveraging these physiological signal properties, this study innovatively employs the high-SNR EMG as a supervisory signal to construct a mapping model from EEG to muscle activation—namely, the elbow joint Brain-Muscle Atlas (BMA-Elbow). The core mechanism of this atlas is as follows: during the training phase, it utilizes synchronously acquired EEG-EMG data to learn the mapping from cortical activity patterns to corresponding muscle activation patterns; during the online control phase, the system requires only EEG input to generate "virtual muscle activity signals" aligned with movement intention via this atlas, thereby driving joint motion. Consequently, the purpose of the Brain-Muscle Atlas is not to replace EMG-based interfaces, but to architecturally compensate for the inherent limitations of non-invasive BCIs—namely, poor raw signal quality and limited movement resolution—by introducing "muscle activation" as a crucial physiological relay layer. This work lays a novel technical foundation for future general-purpose motor decoding systems targeting completely paralyzed patients.

   	As a proof of principle, this study constructed the first brain–muscle atlas for elbow flexion-extension and accordingly designed a brain-muscle-elbow interface. The focus of this work is a methodological breakthrough, aiming to demonstrate the feasibility and advantages of the architecture of "enabling biomimetic signal conversion via the brain–muscle atlas." Experimental results show that this interface can generate continuous and stable elbow joint control commands from low-SNR EEG signals, with smoothness and robustness significantly superior to traditional end-to-end decoding models.
   	
   	The main contributions of this paper are as follows:
   	\begin{enumerate}  
	   	\item This paper implements the first brain–muscle atlas for the elbow joint, transforming the theoretical mapping relationship between the cortex and muscles into a computable, verifiable structured model. This provides the first paradigm for establishing a physiologically consistent decoding framework at the limb level and lays a methodological foundation for constructing future brain–muscle atlas for multi-joint coordination or even whole-body movement.
	   	
	   	\item Based on the atlas, this paper pioneers the brain-muscle-elbow interface, which completely abandons the traditional "direct brain signal control" architecture. By introducing the "muscle activation representation" as a key relay, this interface significantly enhances the naturalness, continuity, and stability of control, opening a novel path for next-generation non-invasive BCIs towards physiological and intelligent control.
	   	
   	\end{enumerate}
   		   	
	   	The structure of the remainder of this paper is as follows: Section \ref{Methodology} details the design and implementation of the proposed method; Section \ref{Result} presents the experimental results; Section \ref{Discussion} discusses the significance of the findings and provides a conclusion.

	\section{Methodology}
    \label{Methodology}
	\subsection{Signal Acquisition}
	
\begin{figure}[h]  
		\centering  
		\includegraphics[width=0.5\textwidth]{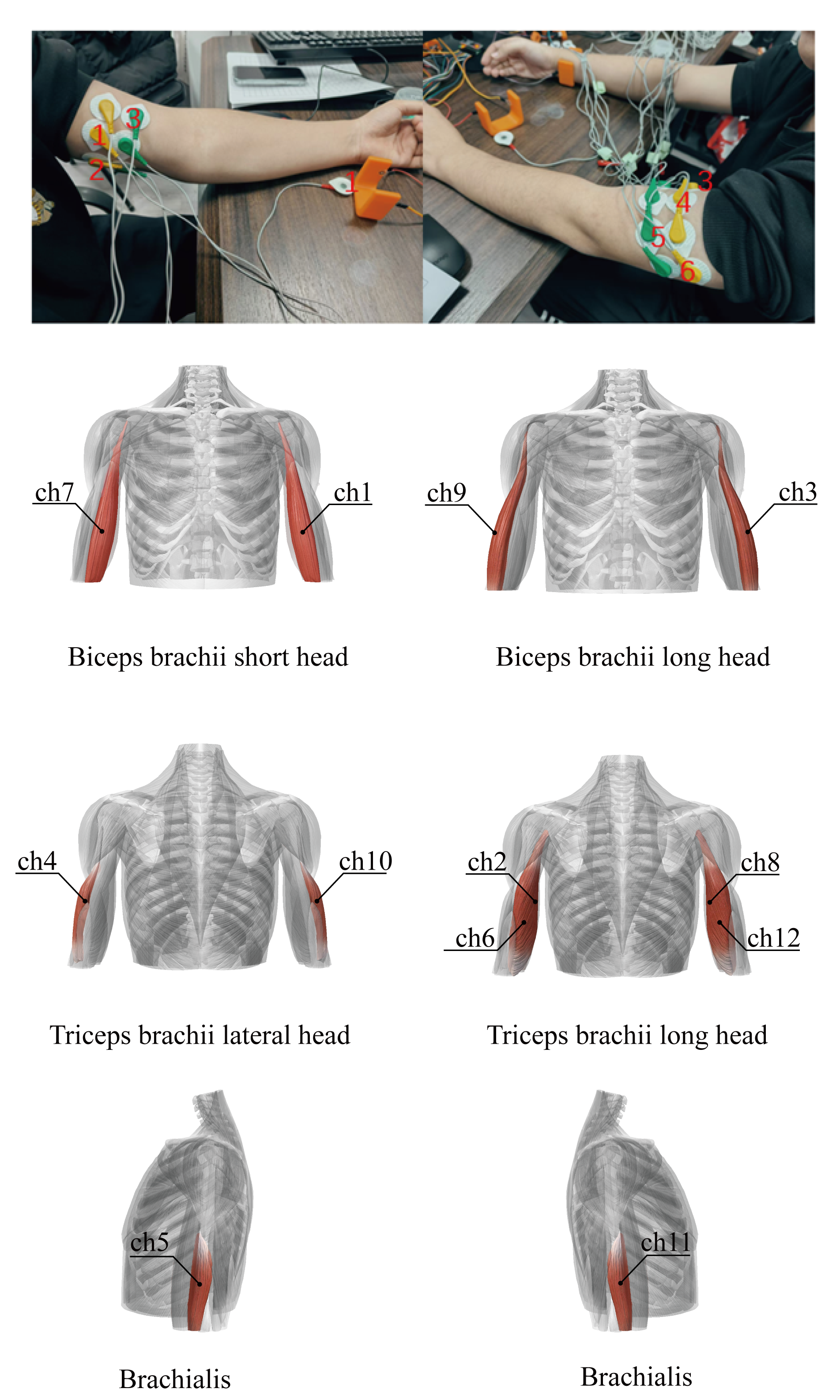} 
		\caption{\textbf{Figure 2.} Electrode distribution} 
		\label{fig:fig2}
	\end{figure}

\begin{figure*}[h]  
		\centering  
		\includegraphics[width=1\textwidth]{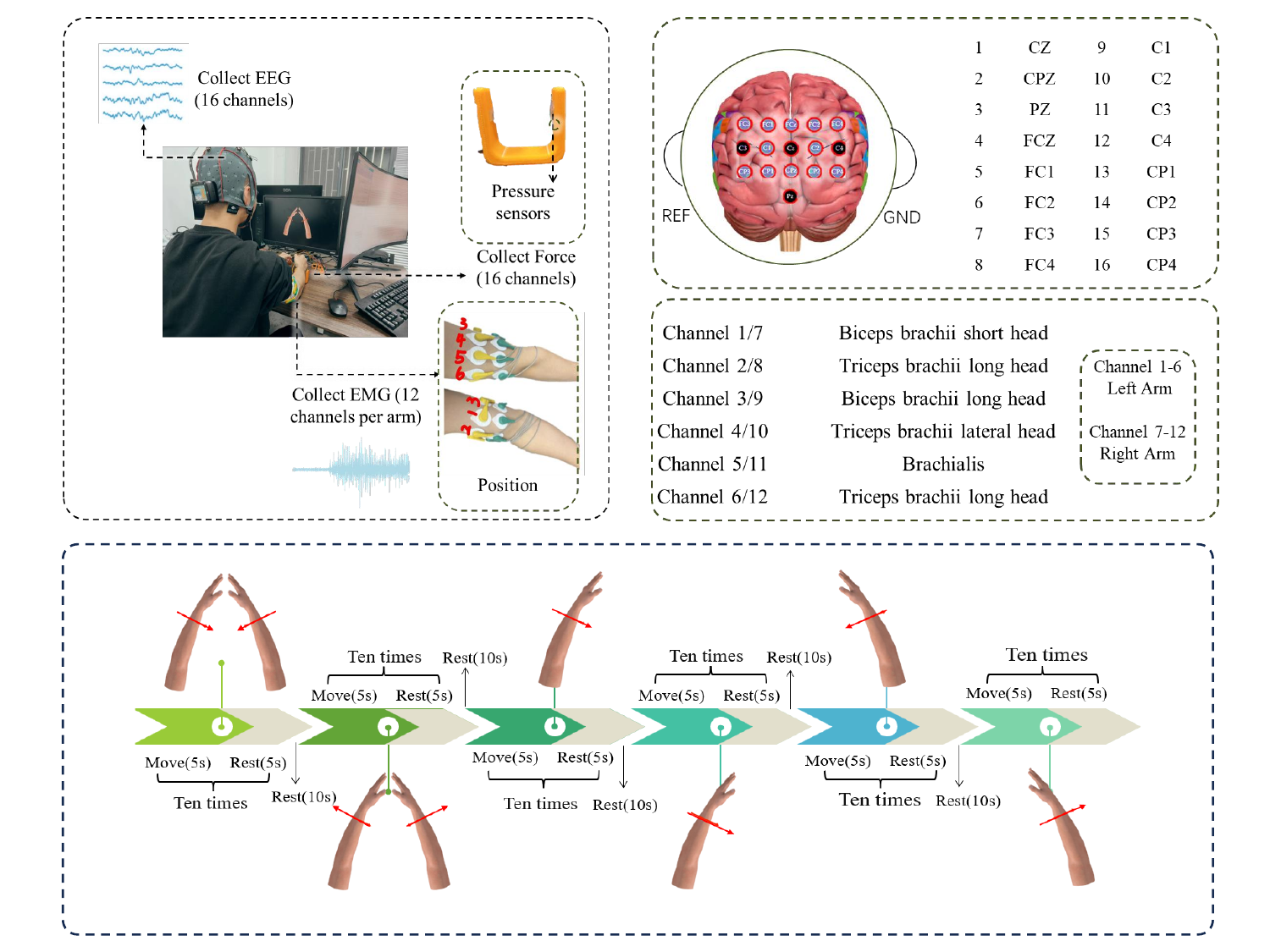} 
		\caption{\textbf{Figure 3.} Brain-muscle interface paradigm. (A)The situation of the subjects at the time of collection and the distribution of electrodes in the arm; (B)Distribution of electrodes for electroencephalogram and electromyography acquisition; (C)Experimental paradigm} 
		\label{fig:fig3}
	\end{figure*}

	A wearable 16-channel non-invasive EEG system (g.Nautilus) was used to record brain activity over the sensorimotor areas of the frontal and parietal lobes at a sampling frequency of 500 Hz. EEG electrode placement strictly followed the international 10-10 system for standardized positioning~\cite{10,11}. The specific spatial distribution of these electrodes is illustrated in \autoref{fig:fig3}. Simultaneously, EMG data were collected using 12 channels, with 6 electrodes evenly placed on each upper arm to cover muscles involved in elbow movement, at a sampling frequency of 1000 Hz, electromyography electrode placement is illustrated in \autoref{fig:fig2}. A pressure sensor was used to record force signals at a sampling frequency of 6.6 Hz (see \autoref{fig:fig3}).
		
	As shown in \autoref{fig:fig3}, during the experiment, participants sat upright in a chair with both hands resting naturally on the table, palms placed and secured within a measurement device. Experimental tasks were presented via a screen positioned directly in front of the participant, who was required to perform specific actions as instructed. Meanwhile, the system simultaneously recorded EEG, EMG from the upper arm, and the exertion of force at the elbow.
    
	\subsection{Signal preprocessing}
	Initially, the EEG, EMG, and force signals were temporally aligned through upsampling procedures. The EEG signals were then pre-processed using a common average reference (CAR) technique and band-pass filtered between 15-35 Hz~\cite{14}. The CAR method involves computing the average signal across all channels to generate a reference, which is subsequently subtracted from each individual channel~\cite{12,13}. The operation is mathematically described as: 
	$$
	x_{i}^{CAR}\left( t \right) =x_i\left( t \right) -\frac{1}{C}\sum_{j=1}^c{x_j}\left( t \right) 
	$$
	where $x_j\left( t \right) $ is the potential value of the $j$ th channel and $C$ is the total number of channels. This processing step effectively removes common mode noise and enhances the spatial resolution of the signal.
	
	For EMG signals, a 20-450 Hz band-pass filter removed the DC offset while retaining muscle activity frequencies~\cite{15,16,17}, followed by a 48-52 Hz Butterworth notch filter to eliminate 50 Hz interference, enhancing signal quality~\cite{18}.
	
	\subsection{Signal Segmentation}
	The EEG, EMG, and force signals were segmented into 60 independent trials. Within each trial, the arm force signal was analyzed to determine the participant's force exertion state. A predefined threshold was established, and any period where the force signal exceeded this threshold was identified as an active task phase. Signal segments meeting this criterion were extracted for subsequent analysis. This segmentation strategy ensured that only task-relevant, high-quality data were retained, thereby enhancing the reliability and validity of the analysis.
	
	\subsection{Construction and application of the Brain-Muscle Atlas}
\begin{figure*}[h]  
		\centering  
		\includegraphics[width=1\textwidth]{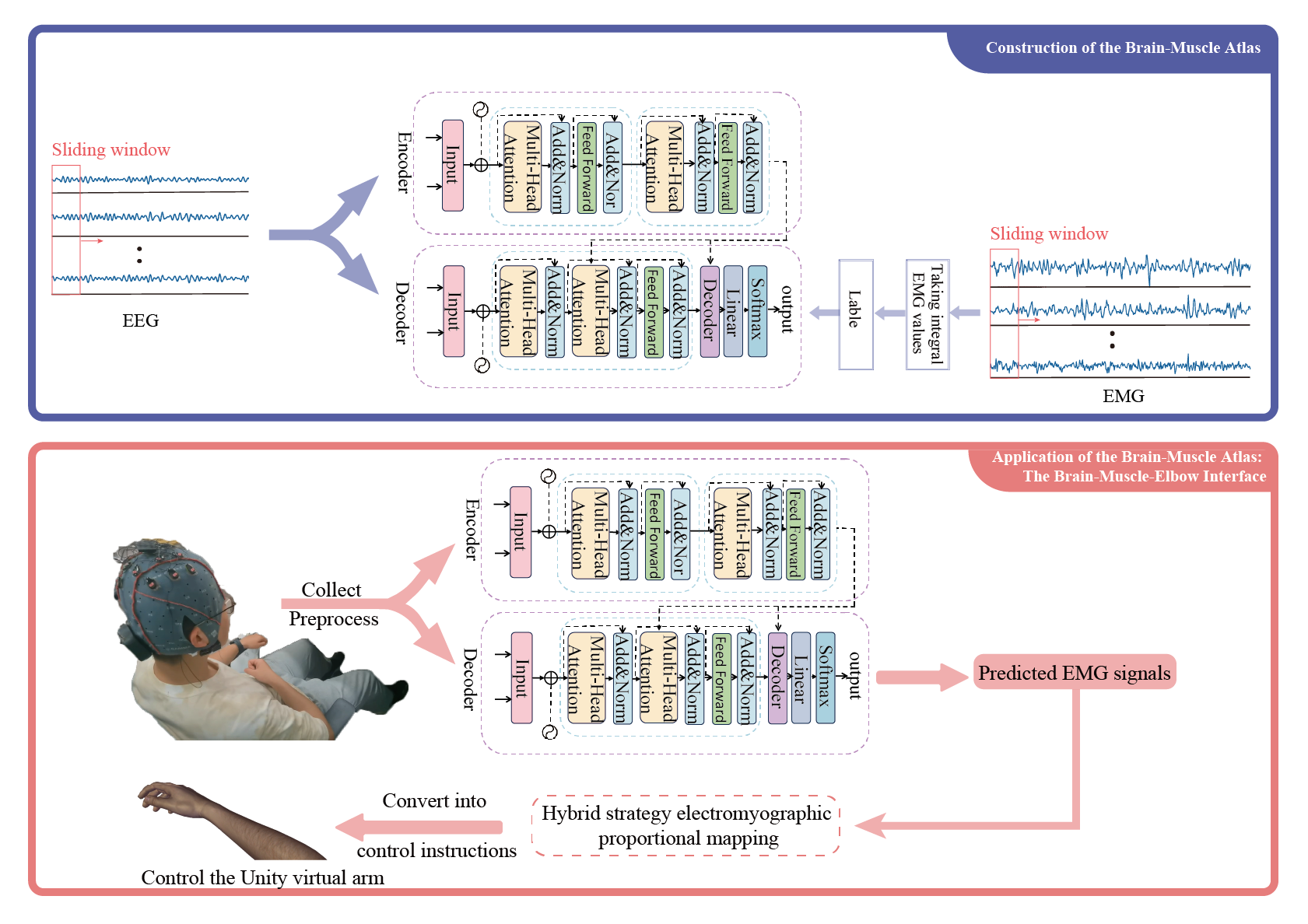} 
		\caption{\textbf{Figure 4.} The framework of Brain-Muscle Atlas} 
		\label{fig:fig4}
	\end{figure*}

	This paper proposes a sliding-window Transformer model for constructing a brain–muscle map (see \autoref{fig:fig4}). Originally developed for natural language processing, the Transformer architecture employs an encoder–decoder framework with multi-head self-attention to capture long-range dependencies and extract coherent global structure~\cite{19,20}. Beyond language, Transformers have demonstrated strong capabilities in modelling temporal sequences and long-timescale dependencies~\cite{21}. Recent studies further show that Transformer-based models are highly effective for neural signal decoding and EEG representation learning~\cite{22}, we extend its use to neural signal decoding, leveraging its strength in modelling long-timescale temporal dependencies to characterise the internal structure of EEG signals within local temporal windows.
	
	To preserve temporal continuity while modelling local dynamics, a sliding-window mechanism is incorporated. This design retains the temporal overlap between adjacent segments, allowing the model to accumulate information progressively and to establish a continuous decoding pathway from cortical activity to muscle activation. In doing so, the approach highlights the Transformer’s adaptability in handling non-stationary, noisy and temporally complex biological signals, thereby broadening its applicability to cross-modal neural decoding tasks.
	
	During construction of the brain–muscle map, synchronously acquired EMG signals serve as supervisory targets, enabling the model to learn explicit cortico–muscular mappings. In its downstream application—the brain–muscle–elbow interface—the system no longer requires EMG inputs; instead, real-time EEG alone is continuously translated into EMG-like representations to achieve natural and responsive control of the virtual hand.

	\subsection{Direction-Proportional EMG Control(DP-EMG Control)}
	
	\begin{figure}[h]  
		\centering  
		\includegraphics[width=0.34\textwidth]{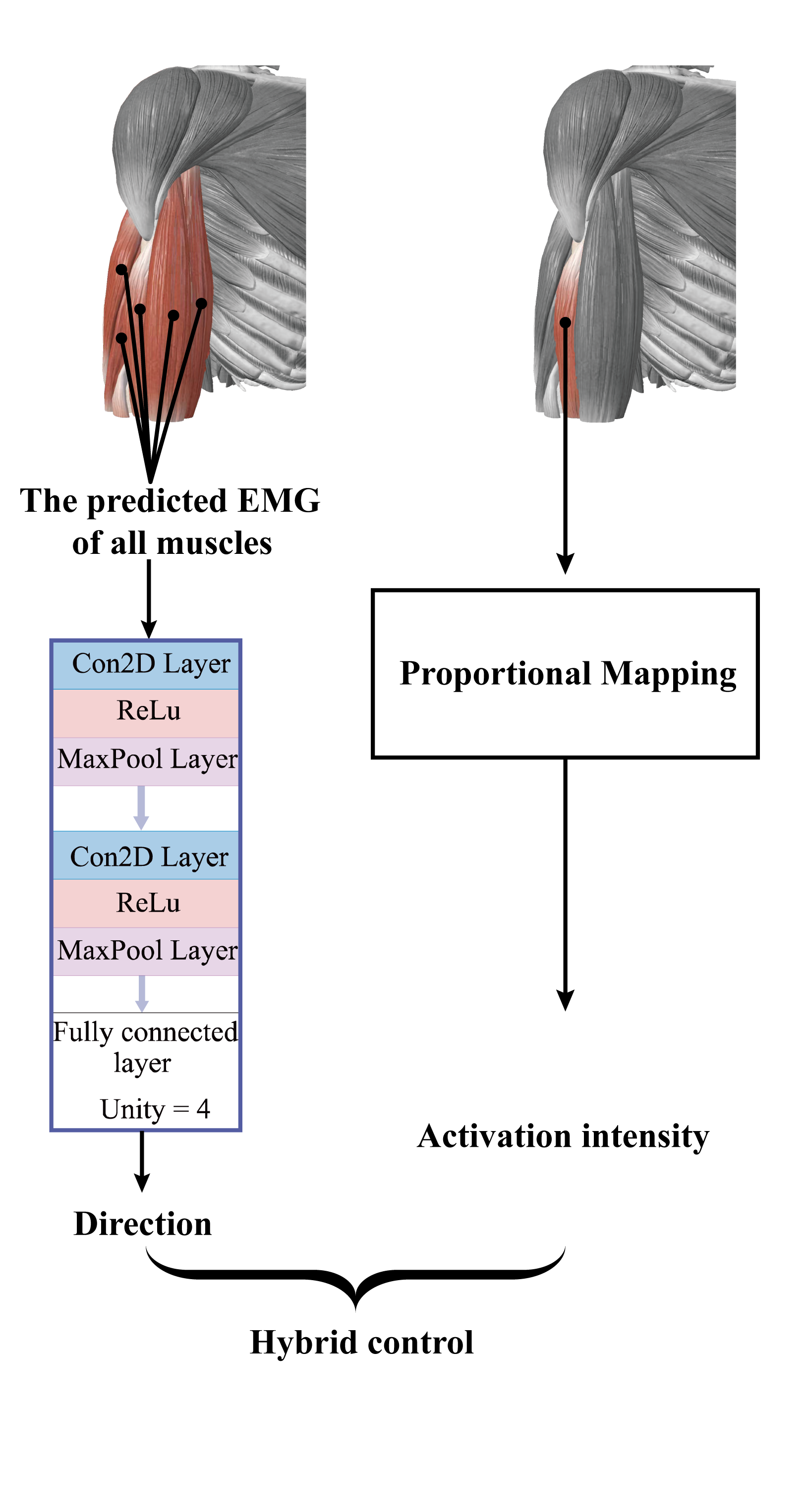} 
		\caption{\textbf{Figure 5.}  Direction-proportional electromyography control method} 
		\label{fig:fig5}
	\end{figure} 
	
	 We propose a hybrid control strategy to realize the application of the brain-muscle mapping(see \autoref{fig:fig5}). First, a convolutional neural network (CNN) is employed to decode the spatiotemporal features of multichannel EMG, producing discrete movement direction commands and thereby improving directional recognition accuracy. Second, the muscle channel with the most reliable prediction performance—typically one associated with elbow flexion or extension—is selected as the source of activation strength. Its EMG activity is used for proportional mapping to realise continuous control. This strategy leverages the CNN’s strength in directional classification while incorporating the natural continuity provided by muscle-activation-based control, thereby achieving a balance between precision and intuitive responsiveness.

	\section{Result}
	\label{Result}
	To systematically validate the effectiveness of this method, this study conducted experiments at two levels: first, the offline predictive performance of the Brain-Muscle Atlas was verified; subsequently, an online experiment involving continuous control of a virtual elbow joint was performed using the trained atlas.
	\subsection{Offline Validation of the Brain-Muscle Atlas}
	\subsubsection{Can the muscle representations translated by the Brain-Muscle Atlas faithfully reflect genuine muscle activity?}
	To systematically assess the feasibility and physiological consistency of the elbow brain–muscle atlas, we conducted an offline validation experiment. A total of 49 healthy right-handed participants with no history of neurological or neuromuscular disorders were recruited. All participants provided written informed consent, and the study protocol was approved by the Ethics Committee of Zhengzhou University and strictly adhered to the Declaration of Helsinki and relevant Chinese ethical regulations.
	
	Following successful construction of the elbow brain–muscle atlas (BMA-Elbow), we used the atlas to translate EEG signals into muscle-activation representations. Specifically, the atlas provides a structured mapping from cortical activity to peripheral muscle activity. Building on this mapping, the brain–muscle–elbow interface learnt to reconstruct EMG patterns associated with movement intention, yielding an “enhanced expression’’ of motor intent.
	
\begin{figure*}[h]  
		\centering  
		\includegraphics[width=0.8\textwidth]{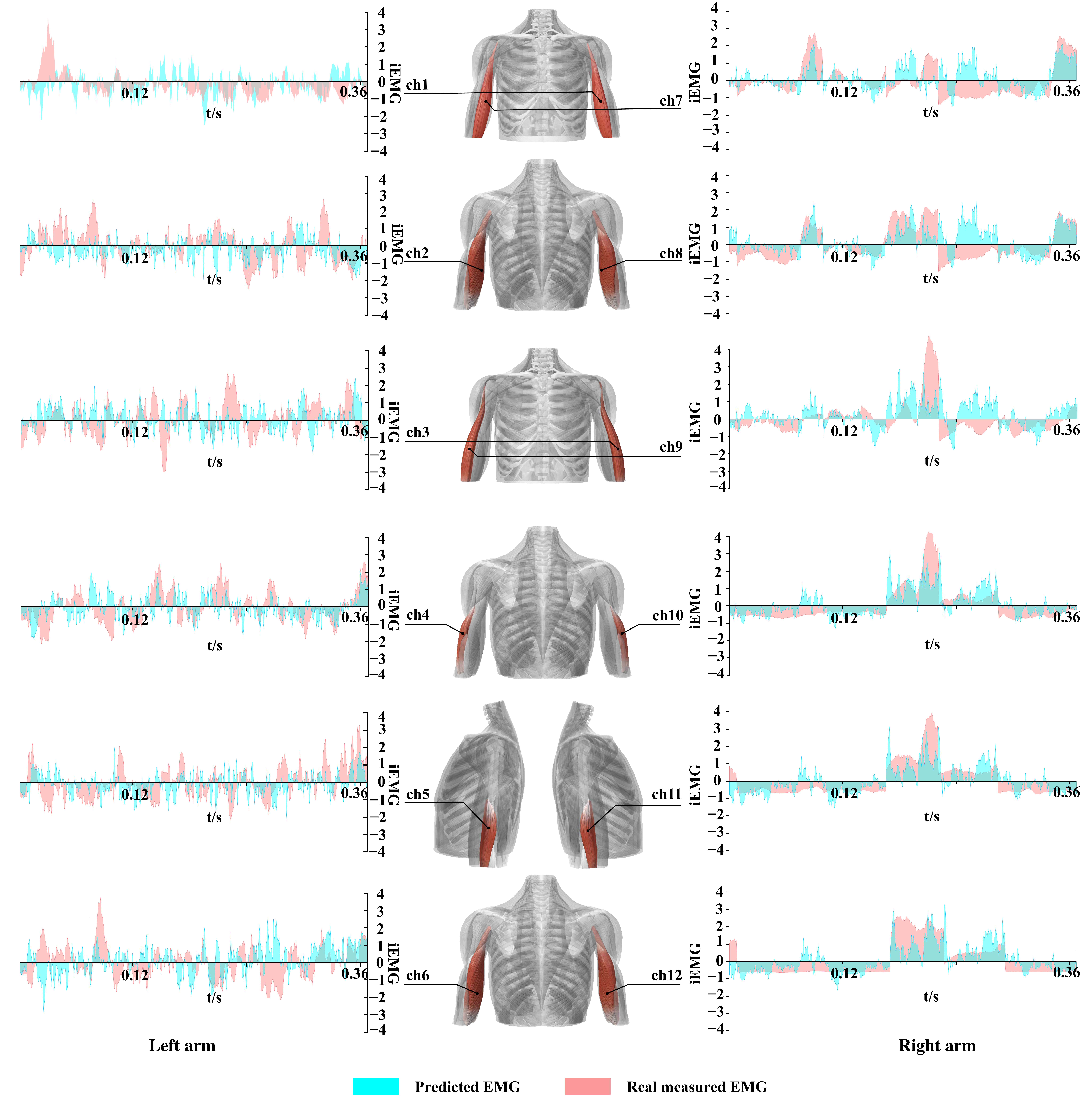} 
		\caption{\textbf{Figure 6.}  Comparison between the EMG representations translated from the brain–muscle atlas and the actually measured EMG signals} 
		\label{fig:fig6}
	\end{figure*}

\begin{figure*}[h]  
		\centering  
		\includegraphics[width=0.8\textwidth]{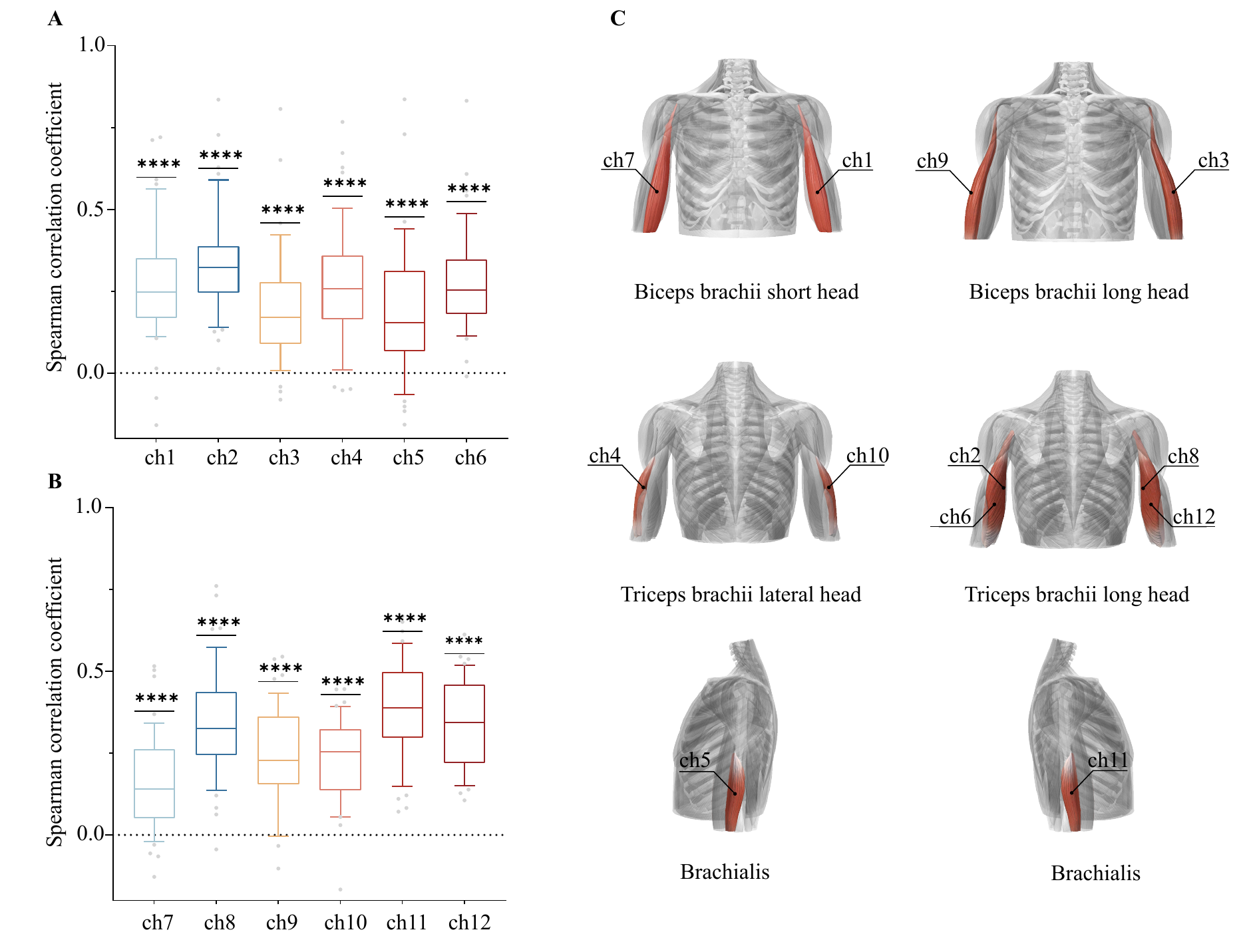} 
		\caption{\textbf{Figure 7.The statistical of the Spearman correlation coefficients between the EMG signals translated by the brain-muscle atlas  and the actual measured EMG values.} (A)results for the left arm;(B)results for the right arm;(C) the distribution of 12 electrodes on the muscle surface} 
		\label{fig:fig7}
	\end{figure*}

	To quantify how accurately the atlas-generated muscle representations captured true muscle activity, we computed the Spearman correlation coefficient (SCC) ~\cite{23} between the atlas outputs and the recorded EMG signals. A one-sample t-test was applied to determine whether the mean correlation coefficient was significantly greater than zero—indicating the presence of a genuine association rather than chance correspondence. \autoref{fig:fig6} presents a representative comparison of atlas-generated and recorded EMG time series for one participant. \autoref{fig:fig7}A and B summarize the SCC distributions across all twelve muscle channels and all participants, while \autoref{fig:fig7}C labels the anatomical locations of the EMG channels.
	
	The results in \autoref{fig:fig6} demonstrate a clear lateralization in reconstruction performance: the reconstructed activity of the right arm closely tracked the overall trend of the measured EMG, whereas the reconstruction for the left arm showed an offset. This pattern accords with the stronger cortical drive, more coordinated muscle synergies, and more efficient sensorimotor integration typically observed on the dominant side during motor execution in right-handed individuals~\cite{24,25,26}. As shown in \autoref{fig:fig7}, the overall mean SCC across all participants and channels was 0.28. Notably, channels 2, 8, 11, and 12 exhibited higher mean SCCs of 0.34, 0.35, 0.39, and 0.33, respectively. These channels correspond to the biceps brachii and triceps brachii—primary agonist–antagonist pairs driving elbow flexion and extension—indicating that the atlas captured the muscle activation states of the key muscle groups particularly well~\cite{27,28,29}. Moreover, one-sample t-tests confirmed that both the mean and maximum PCC values for all twelve channels were significantly above zero ($p < 0.0001$), demonstrating that the atlas-translated EMG signals reflect genuine muscle activity rather than random noise.
	
	Taken together, these results show that the brain–muscle atlas effectively translates EEG signals into muscle-activation representations and reliably reflects true physiological muscle activity.
	
	\subsubsection{Do the representations learnt by the brain–muscle atlas faithfully reflect the physiological pathways of human motor control?}
\begin{figure*}[h]  
		\centering  
		\includegraphics[width=0.8\textwidth]{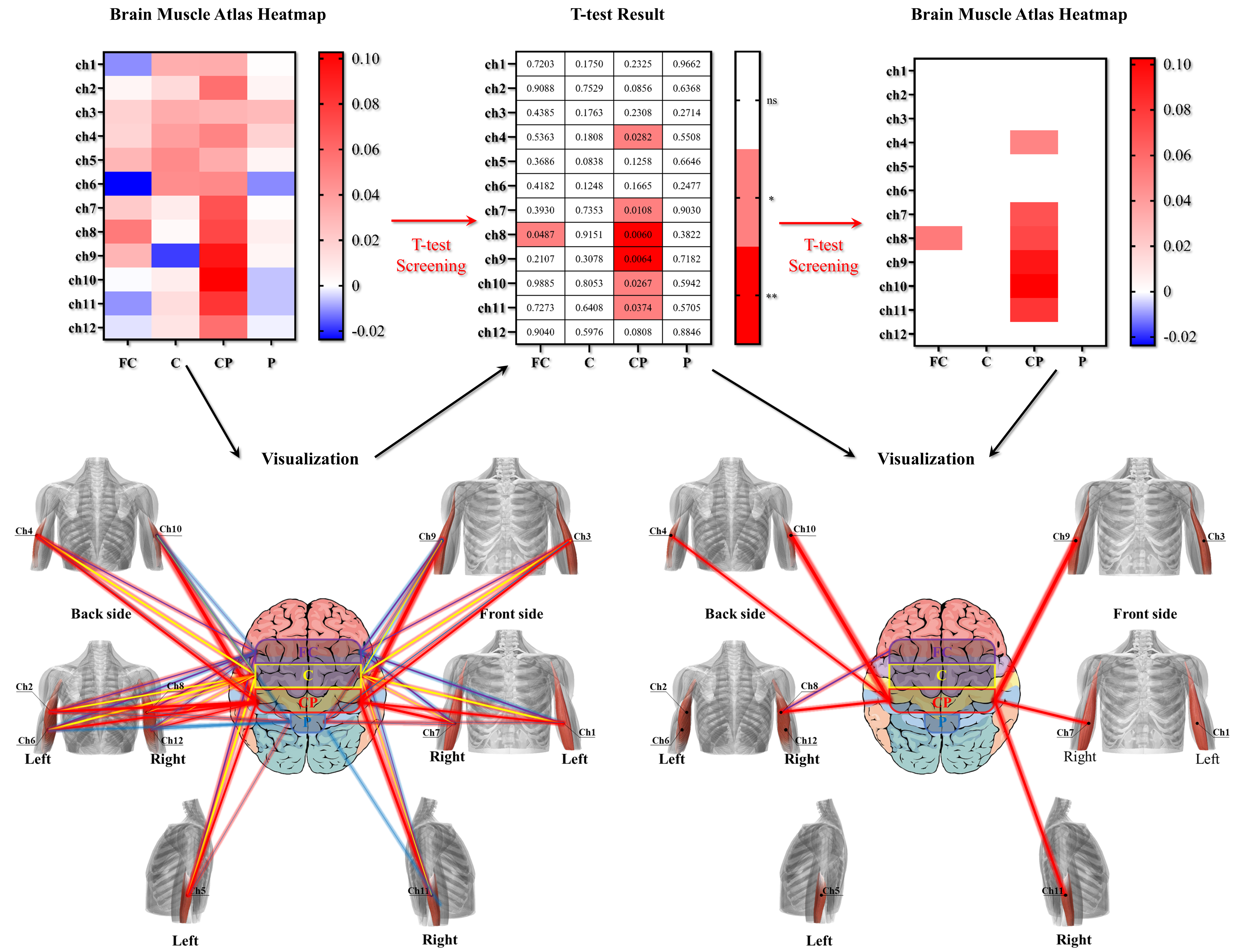} 
		\caption{\textbf{Figure 8. Visualization of Brain Muscle Atlas.} This figure illustrates the topological architecture of the Brain–Muscle Atlas derived from zero-masking perturbation analysis. The panels display the contribution heatmaps and corresponding connectivity pathways before and after statistical screening. Within the Atlas topology, the visual encoding is defined as follows: the line body color identifies the source cortical region to trace the origin of neural drive; line thickness is proportional to the absolute magnitude of the contribution, with thicker lines indicating a stronger functional connection (sensitivity) in the Atlas; and the surrounding halo denotes coupling polarity, where a red glow signifies a positive contribution and a blue glow signifies a negative contribution. By filtering out non-significant connections (one-sample t-test, $p > 0.05$)} 
		\label{fig:fig8}
	\end{figure*}

	To determine whether the representations learned by the brain–muscle atlas genuinely reflect the physiological pathways of human motor control—rather than merely fitting statistical dependencies—we employed a zero-masking perturbation approach to quantify the contribution of each cortical region to sEMG reconstruction. In this procedure, EEG channels associated with individual cortical areas were sequentially masked, and the impact was evaluated by computing the mean squared error (MSE) between the reconstructed sEMG signals obtained before and after masking. Based on the derived contribution matrices, we initially constructed global heatmaps to visualize the weight distribution of brain–muscle couplings. To ensure the statistical reliability of these patterns, one-sample t-tests were conducted across all participants, serving as a screening mechanism wherein connection strengths failing to reach significance were set to zero, thereby retaining only robust, physiologically plausible features. To further elucidate the spatial topology of these interactions, we generated brain–muscle connectivity maps for both pre- and post-screening states using a multi-dimensional visual encoding scheme: the primary color of each connection line identifies its source brain region; line thickness is proportional to the absolute magnitude of the contribution (sensitivity), reflecting the strength of regulatory input; and a surrounding halo indicates polarity, with red and blue glows representing positive and negative contributions, respectively (see \autoref{fig:fig8}). The resulting analysis demonstrated that the centro-parietal (CP) region made a significant contribution to EMG reconstruction, whereas the fronto-central (FC), central (C), and parietal (P) regions did not reach significance.
	
	 This spatial pattern is highly consistent with the neural control mechanisms required by the task: during quasi-isometric contractions against resistance, the brain primarily relies on a sensorimotor integration network centred on the primary motor cortex (M1) and primary somatosensory cortex (S1) for closed-loop regulation~\cite{30,31,32}. The CP region, which spans the M1–S1 boundary, is widely recognised as a key cortical hub for force modulation, utilization of sensory feedback, and proprioceptive integration. In addition, we observed that the fronto-central (FC) region exhibited a significant contribution for channel 8, corresponding to the long head of the biceps brachii. Although this effect was spatially localised, the premotor cortex—located within the FC region—is known to play a critical role in proximal upper-limb control, including postural regulation, synergistic activation, and movement preparation. This finding therefore reasonably reflects the involvement of premotor areas in modulating the long head of the biceps brachii~\cite{33,34}, a primary agonist in elbow flexion, thereby verifying that the brain–muscle atlas captures functionally meaningful mappings between specific cortical territories and muscle activity. Taken together, the cortical contribution patterns revealed in this study are not only statistically robust but also closely aligned with current neurophysiological understanding of sensorimotor control. These results indicate that the representations extracted by the brain–muscle atlas faithfully reflect the cortical functional networks engaged in real human motor control.
	
	Further analysis of the CP-region contribution matrix across individual muscle channels showed a pattern consistent with well-established physiological principles of brain–muscle coordination. For the dominant limb (right arm), the CP region exhibited significant contributions to both the primary agonists (e.g. brachialis, biceps brachii) and their antagonists (triceps brachii). This pattern is consistent with the well-established co-contraction strategy employed during stable force production, whereby the motor cortex regulates both flexors and extensors in concert to enhance joint stability and resistance to perturbation~\cite{35,36}. By contrast, on the non-dominant side (left arm), only the triceps brachii displayed a significant contribution—a pattern that aligns with the common postural compensation observed in bilateral tasks, in which the non-dominant limb primarily provides positional support through isometric extensor activation rather than contributing major driving force~\cite{37,38}.
	
	The observed cortical contribution patterns dovetailed with model performance in the EMG-prediction task: reconstruction accuracy was consistently higher for right-side muscles than for left-side muscles, mirroring the asymmetry in cortical contributions. This concordance suggests that the neural features decoded from EEG faithfully reflect physiologically grounded asymmetries in bilateral motor control. On the basis of these offline findings, subsequent online experiments focused exclusively on real-time control of the right arm.
	
	Taken together, the brain–muscle atlas not only reconstructs motor intention with high fidelity but also exhibits internal cortical-weighting patterns and muscle-specific mappings that closely match established neuroanatomical structures, motor-control theories, and biomechanical principles. The model therefore demonstrates not only strong predictive performance but also clear, verifiable, and mechanistically interpretable physiological rationality, providing a robust scientific foundation for the development of the brain–muscle–elbow interface.
	
	\subsection{Online Control Based on the Brain-Muscle-Elbow Interface}
	\subsubsection{Can the brain–muscle atlas support real-time motor control?}
	To evaluate the practical application potential of the Brain-Muscle Atlas for real-time motion control, we constructed the BMEI based on this atlas and conducted an online control experiment.
	
	For this phase, ten new healthy participants were recruited. First, following the same protocol as the offline experiment, EEG, surface electromyography (sEMG), and force signals were synchronously collected from each participant to train their personalized BMA model. Each participant performed 120 elbow flexion-extension trials. During data collection, participants executed movements according to on-screen cues, while researchers monitored signal quality and movement consistency in real-time to ensure the reliability of the training data.
	
	\begin{table}[ht]
		\centering
		\caption{Time taken by participants to complete the Unity virtual hand online experiment}
		\label{tab:table1}
		\begin{tabular}{ccc}
			\toprule
			\textbf{Participants} & \textbf{stretch the elbow} & \textbf{bend the elbow} \\
			\midrule
			Participant 1 & 7s & 6s \\
			Participant 2 & 5s & 7s \\
			Participant 3 & 8s & 6s \\
			Participant 4 & 23s & 5s \\
			Participant 5 & 5s & 10s \\
			Participant 6 & 5s & 7s \\
			Participant 7 & 41s & 5s \\
			Participant 8 & 18s & 12s \\
			Participant 9 & 5s & 8s \\
			Participant 10 & 7s & 8s \\
			\bottomrule
		\end{tabular}
	\end{table}
	
	\begin{figure*}[h]  
		\centering  
		\includegraphics[width=0.8\textwidth]{fig9} 
		\caption{\textbf{Figure 9.}  Subjects control virtual arm movements} 
		\label{fig:fig9}
	\end{figure*} 
	
	In the subsequent online control stage, participants wore only the EEG acquisition equipment. The system acquired their EEG signals in real-time, decoded them into corresponding "virtual" EMG signals via the individualized BMA, and then transformed these signals into continuous drive commands through the "muscle-to-elbow" conversion module of the BMEI to control a virtual elbow joint model. \autoref{fig:fig9} illustrates this complete pipeline for achieving continuous, real-time, closed-loop control.
	
	Online performance was primarily evaluated using the time required to complete a single flexion-extension task. \autoref{tab:table1} summarizes the task completion times for all ten participants. \textcolor{blue}{Supplementary Videos S1-S10 document the complete real-time control process.}
	
	The experimental results demonstrate that the BMA-based BMEI can stably decode movement intention from real-time EEG signals and generate muscle-level commands sufficient to drive continuous motion of the virtual elbow joint. All participants successfully completed the online control tasks, verifying the feasibility and application potential of the Brain-Muscle Atlas for online brain-controlled applications.
	
	\section{Discussion}
	\label{Discussion}
	
	The BMA and BMEI proposed in this study represent a paradigm shift in motor BCIs. Rather than pursuing end-to-end mappings that bypass biological structure, this approach computationally emulates the natural brain–muscle–joint pathway. The essence of this shift lies in adhering to the hierarchical organisational principles of the motor nervous system. Previous research shows that the central nervous system stabilises and simplifies motor control through multiple intermediate functional layers—such as spinal interneuronal circuits, muscle-synergy modules, and biomechanical regularities intrinsic to the musculoskeletal system~\cite{39,40}. This study is the first to operationalise this principle within a non-invasive motor BCI: by introducing muscle-layer representations as an intermediate decoding space, it translates the neurophysiological concept of muscle synergy into a computational module that can be embedded within the control architecture.
	
	This strategy addresses the intrinsic limitations of end-to-end decoding. In conventional paradigms, direct decoding requires the model to learn both cortical signal interpretation and biomechanical prediction simultaneously, resulting in diffuse and unstable training objectives. In contrast, the present work introduces a physiologically meaningful intermediate representation—muscle activation—which naturally separates the task of cortical signal interpretation from that of dynamical mapping. Under this formulation, the model need only learn the relationship between EEG activity and muscle activation, without directly traversing the complexities of movement dynamics; the predicted muscle representation is then converted into elbow kinematics by the muscle–elbow interface. In doing so, the processes of neural decoding and biomechanical mapping become analytically disentangled, yielding a more stable decoding pipeline that more closely reflects the physiological structure of human motor control.
	
	Offline experiments demonstrated that the muscle representations generated by the brain–muscle atlas accurately captured true EMG activity, with superior reconstruction performance for the dominant limb, where motor control is more finely regulated. This finding not only validates the decoding accuracy but also indicates that the model successfully captures the physiological lateralization inherent to human motor control. Further cortical-contribution analyses revealed that the CP region—implicated in force regulation and sensorimotor integration—made significant contributions to EMG reconstruction. This observation is fully consistent with established neurophysiological theories, which posit that limb movement is regulated through integrated activity across motor and somatosensory cortices, thereby confirming the interpretability of the model~\cite{41,42}.
	
	It is noteworthy that this design philosophy of "embedding a biological intermediate layer" is not only grounded in neuroscience but is also becoming a prevalent trend in high-performance robotic systems. For instance, Tesla's next-generation dexterous hand employs a "tendon-style" force transmission structure instead of traditional direct motor drives. The core purpose is to reconstruct a biological intermediate functional layer akin to "muscle-tendon-joint" to achieve more natural motion trajectories, higher compliance, and better disturbance rejection. The active incorporation of such bio-inspired intermediate layers in engineering systems shares a structural consistency with the BMA and BMEI proposed here: both fundamentally attempt to restore the evolutionarily optimized hierarchical organization in motor control, rather than burdening the decoder or controller with an unnecessary end-to-end task.
	
	This structural convergence across biological and artificial systems reveals an important insight: future human-machine interaction systems that are natural, stable, and high-precision will likely depend on reconstructing the intermediate functional layers shaped by biological evolution within computational frameworks, rather than simply scaling up model sizes or directly approximating complex input-output mappings. The BMA and BMEI demonstrated at the single-joint level in this study provide a computable, verifiable proof-of-concept for this direction.
	
	In summary, this study not only validates performance improvements at the level of elbow joint control but also proposes a generalizable framework that systematically translates the hierarchical control principles of neuroscience into engineering practice. With the future construction of multi-joint and even whole-body Brain-Muscle Atlases, this pathway may establish the core methodological foundation for developing the next generation of motor BCIs that are natural, robust, and physiologically consistent.
	
	\subsection{Conclusion}
	
	This study constructed the first Brain-Muscle Atlas for elbow flexion-extension and developed a Brain-Muscle-Elbow Interface based on it, establishing a hierarchical decoding framework consistent with neurophysiological mechanisms. The core innovation of this framework lies in introducing the "muscle activation representation" as an intermediate level for motor decoding. This shifts the previously bridging "brain to joint" direct mapping back to the biologically grounded "brain to muscle to joint" hierarchical transmission pathway. This design fundamentally avoids the need for direct inference of complex limb dynamics, significantly enhancing the stability of the decoding process and the physiological plausibility of the output commands.
	
	Offline experimental results demonstrate that the constructed Brain-Muscle Atlas can effectively recover temporal dynamic patterns from EEG signals that are highly consistent with actual elbow muscle activation, verifying its capability to accurately capture cortex-muscle mapping characteristics. Online control experiments further confirm that the BMEI, based on this atlas, can translate EEG signals into continuous and stable control commands for elbow joint motion in real-time, enabling natural flexion-extension control of a virtual elbow.
	
	These findings collectively indicate that reconstructing the key relay in the motor neural pathway through the Brain-Muscle Atlas can provide a more robust and interpretable control strategy for non-invasive BCIs. The methodological framework established in this study not only validates the feasibility of achieving physiologically consistent decoding at the single-joint level but also lays a methodological foundation for future construction of multi-joint coordinated and whole-body movement Brain-Muscle Atlas systems.

	\bibliographystyle{elsarticle-num} 
	\bibliography{sss1}

\end{document}